\documentclass[preprint]{elsarticle}

\usepackage{graphicx} 
\usepackage{dcolumn}
\usepackage{bm} 
\journal{Journal of Magnetism and Magnetic Materials}
\begin{document}
\begin{frontmatter}
\title{Late stage, non-equilibrium dynamics in the dipolar Ising model }

\author[tom]{Tom Hosiawa}
\author[macisaac]{A.B. MacIsaac\corref{cor1}} 
\cortext[cor1]{Corresponding author.}
\ead{allanb@uwo.ca}
\address{The Department of Applied Mathematics\\
The University of Western Ontario, London, Ontario, CANADA N6A 5B9}

\begin{abstract}

Magnetic domain structures are a fascinating area of study with interest deriving both from technological applications and fundamental scientific questions. The nature of the striped magnetic phases observed in ultra-thin films is one such intriguing system. The non-equilibrium dynamics of such systems as they evolve toward equilibrium has only recently become an area of interest and previous work on model systems showed evidence of complex, slow dynamics with glass-like properties as the stripes order mesoscopically.  To aid in the characterization of the observed phases and the nature of the transitions observed in model systems we
have developed an efficient method for identifying clusters or  domains in the spin system, where the clusters are based on the stripe orientation.  Thus we are able to track the growth and decay of such clusters of stripes in a Monte Carlo simulation and observe directly the nature of the slow dynamics. We have applied this method to consider the growth and decay of ordered domains after a quench from
a saturated magnetic state to temperatures near and well below the critical  temperature in the two dimensional dipolar Ising model.  We discuss our method of identifying stripe domains or clusters of stripes within this model and present the results of our investigations.

\end{abstract}


\end{frontmatter}

\section{Introduction}

The properties of ultra-thin magnetic thin films have been the subject of study
for some time, but new experimental techniques to create and
probe such systems has intensified the efforts of researchers to
understand these systems in recent years.\cite{prinz1998,heinrich2004,shen2004}  Among the
more interesting properties is the existence of stripe phases. These phases are 
the result of the competition between short-range exchange interactions and long-range,
dipole-dipole interactions.   Such pattern
formation is of great interest as the foundations of future potential magnetic storage
devices are based in part on the properties of such materials.\cite{wolf2001}
However, the phenomena is much more general and similar patterns have
been observed in such diverse systems as Langmuir monolayers, diblock
co-polymers, and type 1 superconductors.\cite{seul1995-a}
More recently similar competition between long-range,
dipole-dipole and short-range interactions has been seen in model
systems as an essential control mechanism in the self assembly
process for nano-structures.\cite{glotzer2007}

The static, equilibrium properties of ultra-thin magnetic films have
been well studied experimentally \cite{heinrich2004} and 
the dipolar Ising model has been considered extensively in efforts to
explain these properties.\cite{debell2000} However a better understanding of the 
stability and dynamics of the domain structures in these materials is still 
needed and is highly desirable given that they determine many of the
technologically important properties of these materials. 
It is widely known that the static and dynamic properties of a material 
can sometimes be determined by clusters or domains within the material. Such is the case the 
two-dimensional Ising
model with short range interactions only\cite{demeo90-a}, where the explicit connection between the
geometric clusters of spins in the same state to the physical clusters which are characteristic of the critical phenomena in the system,  has been formally
established.\cite{coniglio80,coniglio90}  This connection has been exploited to gain
a better understanding of the dynamics of the short range interaction Ising spin system, 
the nature of the phase transition, 
and also to develop acceleration algorithms for Monte Carlo simulations, which have
greatly expanded the utility of the model.\cite{swendsen1987,wolff1989} A
similar connection has yet to be established for the dipolar Ising
model. Given  the potential utility of the dipolar Ising model, and given the very slow 
dynamics that have observed in simulations to date, such a connection
and the development of acceleration algorithms would be quite useful.
The methods and results presented in this work are an initial 
step toward the development of such an algorithm and at the same time
provide insight in to the dynamics of the dipolar Ising model itself.

\subsection{ The dipolar Ising model }

The Hamiltonian for the dipolar Ising model in reduced units can be written as
\begin{equation}
{\cal{H}} = -J\sum_{<ij>} \sigma_{i}\sigma_{j} + 
g \sum_{i,j} \sigma_{i} \Gamma_{ij}\sigma_{j} +
H\sum_{i} \sigma_{i}
\end{equation}
where $<i,j>$ in the first term indicates a sum over all pairs of nearest 
neighbor spins
$i,j$,  $\sigma_{i}= \pm 1$ is the magnetic
moment at site $i$  and $J$ is the strength of the exchange
interaction.  The second term is a long-range dipole-dipole 
interaction of strength $g$ and the sum is over all pairs of spins. The final term treats 
the interaction with an applied field, $H$, which we do not consider in this work. 
The system is assumed to be infinite, but restricted to states which are periodic.  Periodic boundary conditions are therefore used to treat the exchange interaction and Ewald sums are used to account for the long-range nature of the dipolar interaction.  The exact form of $\Gamma_{ij}$ and other details related
to the Ewald summation are as given by Ref.~\cite{booth1995}.
In this paper we assume the spins are perpendicular to the plane of the film. 

\subsection{Previous work}

The equilibrium properties of dipolar Ising model
 have been studied extensively and a 
review of this work
can be found in De'Bell et al. \cite{debell2000} and the works cited
therein.  In the work presented here the ratio of $J$ to $g$ has been fixed  at
$J/g=8.9$, which has been shown previously to lead to an ordered
ground state, in a monolayer, of stripes of width 8 spins parallel to a
lattice axis.\cite{mac95}  The nature of this transition is 
still the subject of some debate, much of which centers on the
characteristics of the equilibrium phase just above the critical 
temperature and the comparison to experimental results.  

In contrast the non-equilibrium dynamics of the dipolar Ising model
have not been so extensively considered.  The earliest work considered
the relaxation of the magnetization from the saturated magnetic state, 
($\{\sigma_{i}\} =1$),
after a quench toward equilibrium at a temperature below $T_{c}$.  
Sampaio, de Albuquerque and de
Menezes\cite{sampaio1996} found that depending on the relative
strengths of the dipolar and exchange interactions, one can have either 
exponential or power-law relaxation.  The time frame considered was very 
short, typically on the order of 5000 Monte Carlo steps (MCS)  at most, and
the relative strength of the interactions was
such that the stripe width in the ground state was limited to at most 4
spins.  Rappoport et al.\cite{rappoport1998} reported similar results
on larger lattices,  but with a truncated dipolar interaction. 

The phenomena of aging in the dipolar Ising model was considered in a number of
articles, all of  which considered small stripes of width one or two spins. By
measuring the spin auto-correlation function,
\begin{equation}
C( t, t_{w} ) = \frac{1}{N} \sum_{i} \left< \sigma_{i}(t+t_{w})
\sigma_{i}(t_{w}) \right>,
\end{equation}
Toloza, Tamarit and Cannas \cite{toloza1998} were able to observe
aging, the nature of which was dependent on the  relative
strength of the interactions.  Stariolo and Cannas\cite{stariolo1999}
expanded upon this earlier work to show how $C( t, t_{w} ) $ crosses
over from logarithmic decay to algebraic decay as the ground state
stripe width increases. Gleiser et al. \cite{gleiser2003}  and Gleiser 
and Montemurro \cite{gleiser2006} measured the time evolution of the 
domain size of stripes of width one and two spins 
and hypothesized that the differences in the nature of the dynamics was 
related to the existence of metastable states.  All of these studies
were concerned with short to intermediate time scales (up to
$t=10^{6}$) and only for the case of small stripe widths.

Bromley  and his co-authors\cite{Bromley2000,bromley2003} considered the 
dynamics of systems with a larger stripe width (8 spins) during a quench 
from the saturated magnetic state toward  equilibrium near the critical  
temperature.  They were able to establish the existence and character
of three relaxation regimes.  They considered results from simulations on
 the order of at most 1,000,000 Monte Carlo steps
(MCS), to develop a general understanding of the relaxation in their
system in each of three regimes that they identified.  
At very early times they found that the dynamics of the system  was characterized by the nucleation 
of small islands, where the spins are oriented in the direction opposite to the initial
 saturation direction. As these small islands grew the net magnetization approached
zero.  In the second regime there remained a small remnant magnetization. This remnant magnetization decays at a rate which is much slower than that in the first regime.  In the final regime, at late times, they found that 
the islands become elongated and arranged into regions which manifest the stripe pattern of the expected 
equilibrium phase.  However these local regions did not yet
share a common orientation and hence the system was not completely in the smectic phase. To support 
this conjecture, they provided
three snapshots of an extended simulation that showed single spin
configurations at times $t=300,000$, $t=600,000$ and $t=900,000$,
where by hand they determined the regions of local smectic
ordering.  Their results were consistent with those of Desai and Roland and Sagui and Desai who had conducted similar studies 
using Langevin dynamics for a model with a continuous uniaxial ``spin'' variable, with an additional  significant difference being the use of open boundaries.\cite{roland1990,sagui1994}

Mu and Ma \cite{mu2002} considered a dipolar model system
under going a quench from a random state.  They discussed qualitatively
the early stages of the relaxation to equilibrium, and felt that the
depth of the quench fundamentally changed the nature of the relation
process.  Their simulations were run for times on the order
of 100,000 MCS for stripes of various width from approximately 1 to 9
spins.  For a deep quench they
claimed that the labyrinth structure they observed became frozen due
to the frustration induced by the competition between the short- and
long-range interactions.

It is the late stage relaxation identified by Bromley and  
Mu and Ma, but which was not considered in depth due to computer time constraints, 
 that we wish to address in this article. We also wish to
consider ground state stripes of the size treated by Bromley and by Mu and Ma, rather than
the very small stripes considered in the other works discussed above.
We have extended the length of simulation by at least a factor of 10 compared
to previous studies, and have
recorded the spin configurations every 1000 MCS.\cite{websitemovies}
Our simulations have a run length of
between 2 million and 14 million Monte Carlo time steps, depending on
the temperature of the simulation. The longest of our simulations
required over 30 days to run using a 4 way parallel code on 4
CPU's.  Thus this single simulation required 120 days of CPU time.  Because of
this huge computational demand we have only 
considered a few temperatures and conducted two sets of simulations at
each temperature.  However even this small number of simulations required
over 4 years of CPU time.

While we will be drawing analogies between the standard 2D Ising
model and the dipolar Ising model,
one must remember that  unlike the standard 2D Ising
model, the ordering observed in the dipolar Ising model is a mesoscopic
ordering of the stripes and not a microscopic ordering of the spins
themselves.  Therefore one must be cautious about characterizing the
transition in terms of spin properties rather than in terms of
properties of the stripes.   For stripes of small width this can be a
serious concern as the spins fluctuations and the stripe fluctuations
can occur on similar length scales.  For example, if one were to
consider a system with stripes of width h=2, a spin flip would be
equivalent to half the width of a stripe and even in the ground state
of the system 100\% of the spins would lie on a stripe boundary.  If
instead one where to consider a system with a ground state of stripes
of width h=8 (or greater), then only 25\% of spins lie on a boundary in the
ground state and a single spin flip is no longer comparable to the
excitation required to disorder the mesoscopic stripe order.  This is
also important as the  smallest of ground state stripe widths seen 
experimentally is still quite large, typically on the order 
of 100-1000 spins.\cite{bauer2004,vaz2008}  Thus $h=8$ must be considered small when comparing
to experimental system, however it is hoped that
scaling arguments can be developed which will allow meaningful comparisons to be made.\cite{whitehead2008}

To study the dynamics and more precisely the late stage relaxation of 
this system we need to be able to define
domains within the system which are based, not on the state of the
spins, but on the local orientation of the stripes.  It is important to note the distinction between clusters of spins or spin domains and  clusters of stripes  or stripe domains. A stripe is in fact a geographical domain within which the spins are in the same state and therefore a stripe may be referred to as a spin domain or a cluster of spins. A stripe domain, or equivalently a cluster of stripes, will be a geographical domain in a striped system within which the stripes share an orientation. It will be the evolution of these stripe domains or clusters of stripes which should define the dynamics of our system. 

The rest of this paper proceeds as follows: 
Section \ref{sec:clusters} discusses
the method used to define the clusters of stripes in the model system and provides
some sample configurations. Section \ref{sec:order} discusses the results
of our simulations in terms of the time evolution of the magnetization and 
the order parameter. Section
\ref{sec:count} reanalyzes the same simulations by measuring the properties of 
stripe domains defined using our method of section \ref{sec:clusters}.  Finally section \ref{sec:concl} summarizes the results.

\section{Identifying clusters by stripe orientation\label{sec:clusters}}

To achieve the goals of this work we need to be able to identify,
within a given spin configuration, domains where the stripes have a particular orientation,
 along either the $x$-axis or the $y$-axis of the underlying lattice.  We also need the
algorithm to be simple and efficient.  The standard methods for
such an analysis are based on the 
work of M. Seul, L. O'Gorman and M. Sammon \cite{seulbook}, which were developed
 in part to support their research related to pattern formation in
numerous experimental systems\cite{seul1995-a} including some novel magnetic
systems.\cite{seul1992a,seul1992b}   We have implemented and used some of those routines
to aid in our analysis. However those routines are quite general and we have developed
efficient routines specifically for our purposes. In future 
work we will systematically compare our results to those that can be obtained 
using the routines of Seul, O'Gorman and Sammon.

\begin{figure}[ht]
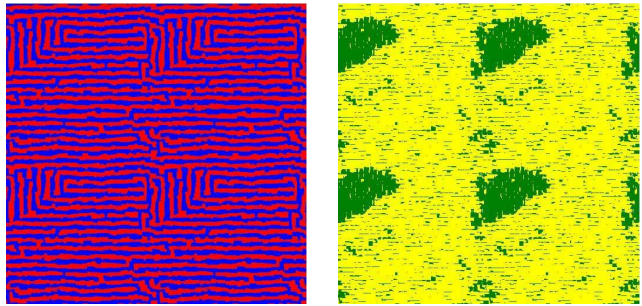

\begin{center}
\begin{tabular}{c c}
\includegraphics[height=40mm]{time=0000500000__4.900-Spin.eps3} &
\includegraphics[height=40mm]{time=0000500000__4.900-Phase-nofilter.eps3}
\\
\end{tabular}
\caption{ (a) Original spin configuration from a $256 \times 256 $
simulation tiled to provide a $512 \times 512$ image. Red(gray):$\sigma=+1$,
Blue(black):$\sigma=-1$
(b) The result of applying the first stage of our cluster finding
method. Green (black):local vertical stripes. Yellow(gray):local horizontal stripes.\label{fig:example} }
\end{center}
\end{figure}

As a first step in our method we will make use of a simple line-of-sight 
argument. For each spin in
the system assume that one is sitting at the spin location in the
lattice.  Now  look in each of the four directions along the lattice
axes ($+x$, $-x$, $+y$ and $-y$) and determine how far in each direction one
would walk before encountering a spin in the opposite state.  Now add
the length one finds in the $+x$ and  $-x$ directions together and the
length one finds in the $+y$ and $-y$ directions together.  If the former
length is larger than the latter length ( $x$ length is larger than $y$
length) then assume that spin location is part of a region with stripes 
oriented along the $x$-axis.  If the opposite is true ($y$ length is larger than
 $x$ length) then one assumes the spin is located in a region with
stripes oriented along the $y$-axis.  Thus one can assign to each point
on the lattice a binary variable which designates the orientation of
the stripes at that point on the lattice.  For example, Figure 
\ref{fig:example}a shows a spin configuration taken from a
simulation just below the critical temperature, but before the system has
reached equilibrium.  The system size is $256 \times 256$, but to
highlight the nature of the stripe domains given the periodic boundary
conditions,  we have tiled 4 copies of the
system together to give a $512 \times 512$ picture.  In this Figure red (gray)
indicates spins with $\sigma = +1 $ and blue (black) indicates spins with
$\sigma = -1$.  Figure \ref{fig:example}b shows how
our the line of sight algorithm would classify each point in the
system.  In this Figure green (black) represents points where the systems
appears to be ordered vertically (along the y-axis) and yellow (gray)
represents horizontally ordered regions (along the x-axis).

\begin{figure}[htb]
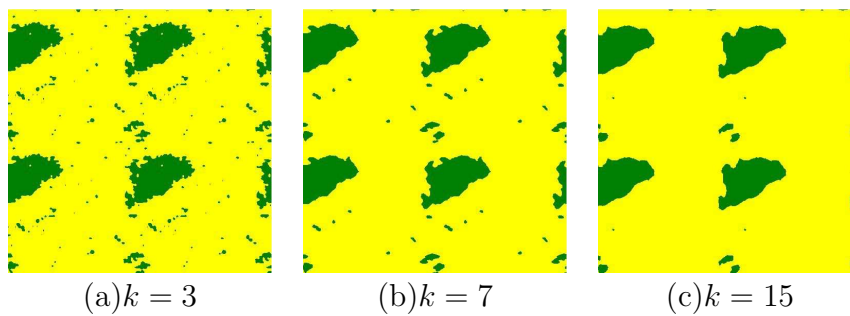

\begin{center}
\begin{tabular}{c c c}
\includegraphics[height=35mm]{time=0000500000__4.900-Phase-medfilter_k=5.eps3} &
\includegraphics[height=35mm]{time=0000500000__4.900-Phase-medfilter_k=9.eps3} &
\includegraphics[height=35mm]{time=0000500000__4.900-Phase-medfilter_k=15.eps3}\\
\large{(a)$k=3$} & 
\large{(b)$k=7$} & 
\large{(c)$k=15$}\\\\ 
\end{tabular}
\caption{ \label{fig:median}
The result of applying a median filter to the image shown in Figure
\ref{fig:example}b with $k$ equal to (a) 3, (b) 7 and (c) 15.
Green (black):local vertical stripes. Yellow(gray):local horizontal stripes. }
\end{center}
\end{figure}

\begin{figure}[htb]
\begin{center}
\includegraphics[height=55mm]{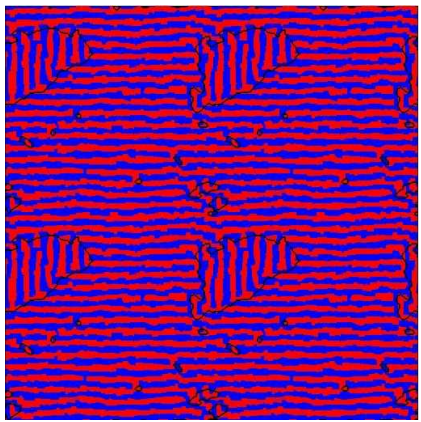}\\
\caption{\label{fig:cool}
The clusters of horizontal and vertical stripes in Figure
\ref{fig:example}a defined after the application of a median filter
using $k$ equal  7 as in Figure \ref{fig:median}b.  The black lines divide regions 
with horizontal and vertical stripe orientations. }
\end{center}
\end{figure}

It is quite evident from Figure \ref{fig:example}b that our routine does 
a reasonable job determining the stripe orientation however, spin
fluctuations can lead to patchiness in the classification. 
 These artifacts are difficult to avoid and are
unimportant when one is concerned with the nature of the ordering in the system, as they reflect microscopic fluctuations and we are
concerned with a mesoscopic scale ordering. 
To provide a clearer picture of the ordered
regions we therefore employ a median filter to our initial classified
image array.  The purpose of such a filter is to reduce speckle
noise,  and functions by replacing the value at a site by the median
value in a $k \times k$ neighborhood centered on that site.  As we
allow only two states in our image, the median filter reduces to a
``majority rules'' calculation.  The choice of $k$ is the only free
parameter in this filter and we show in Figure \ref{fig:median} three  
examples of this filter applied to our initial classified image.  In
Figure \ref{fig:median}a-c we have taken $k=3$, $k=7$, and $k=15$
which are equal to just below one half of a stripe width, just below
one stripe width, and just below than two stripe widths. In Figure
\ref{fig:cool} we have superimposed the boundaries of the stripe domains
shown in \ref{fig:median}b on to the original spin configuration.
As one can see the median filter has the desired effect of reducing
speckle noise  and does not effect the underlying
classification in any significant manner.  For the purposes of this
work we have chosen to use $k=7$, which, as stated above, is just below
the stripe width in the ground state.  Thus small spin-level 
fluctuations do not influence our results, while larger stripe-level 
fluctuations will.

\section{ Magnetization and Order Parameter \label{sec:order} }

\subsection{Magnetization}

We wish to study the late stage dynamics of a quench from the saturated magnetic state, $M=1 $, to equilibrium at temperatures  below the critical temperature in the dipolar Ising model. The early time dynamics
have been considered previously by a number of authors as discussed above.  

As stated above for our chosen ratio the ground state stripe width is $h=8$ spins.  Most previous work 
considered systems where the ground state was stripes of width 1 or 2 spins.
The advantage of choosing such small stripes was that the dynamics, while still much slower than what 
is typical for the 2D Ising model, are still accessible within a
reasonable length simulation.   We do 
not expect our results for wider stripes to necessarily be comparable to these
studies of very narrow stripes, but instead hope they will be
qualitatively similar to experimental results.
It has been shown that the model system we have chosen has a critical
temperature of approximately $T_c = 5.0 \pm 0.1$.  Therefore we have
chosen to simulate quenches below $T_c$ at $T= 4.00, 4.25, 4.50, 4.75$ and $4.90$,
at $T=T_c=5.00$ and just above $T_c$ at $T= 5.10$, and $5.20$.
The variation of the magnetization as a function of time has been well
studied by Bromley et al. for exactly the system considered in this
work.   They showed that for all the temperatures they considered the
magnetization decayed to within \%1 of the saturated magnetization
after about 2000 MCS, however they also showed that some memory of the
original saturated state remains for a much longer time period.  
Our results, at all temperatures considered, are in agreement with those of Bromley et al., and show that the 
magnetization decays effectively to zero within the first 2000 MCS.

\subsection{ Order Parameter }
As pointed out by Bromley et al. the system has not reached equilibrium, within the time frame they considered.
They were able to conclude this because the system should be in a smectic striped phase, but direct observation of the spin configurations clearly showed a
system in the tetragonal phase.  Bromley et al. also used properties of the
magnetization to show that the system had not yet reached equilibrium.  
We are able to show this more directly and quantitatively by 
calculating the orientationial order of the system. To do so we calculate the orientational order 
parameter, $O_{hv}(t)$, defined by Booth et al.\cite{booth1995} as 
\begin{equation}
O_{hv} = \frac{n_{h} -n_{v}}{n_h + n_v},
\end{equation}
where $n_h$ and $n_v$ are the number of bonds between spins with
opposite orientation which are
parallel to the horizontal or vertical direction, respectively.
We have also calculated an alternative, more general, order parameter,
recently proposed by Tan and MacIsaac\cite{tan2007} based on the structure 
factor.  For our model system   in equilibrium below  $T_{c}$ there are two main peaks
in the structure factor, $S(\vec{q})$.  These peaks are along one set of the axes in q-space,
characteristic of stripes parallel to one of the axes in real space;  either 
$x=0$ or $y=0$.   Above 
$T_c$ there are four main peaks in $S(\vec{q})$, as one has stripes along both
axes in real space. If one defines $\vec{q}_+$ and $\vec{q}_-$ as the 
location of the main peaks in q-space associated with 
stripes along $x=0$ and $y=0$ respectively, then one can define a
general order parameter,
\begin{equation}
O_{\pm} =  \frac{ S(\vec{q}_+) - S(\vec{q}_-) }{
S(\vec{q}_+) + S(\vec{q}_-)} ,
\end{equation}
which will be non-zero below $T_{c}$ with a saturation value of plus or minus one
and zero above.  This order parameter has the benefit of being more
general that $O_{hv}$ as it can be defined even in the case where the
stripes are not parallel to a lattice axis. Note that we have measured 
$|O_{\pm}|$ and $|O_{hv}|$ in our simulations, which is appropriate in a finite
system.\cite{binder} 

\begin{figure}
\begin{center}
\includegraphics[height=45mm]{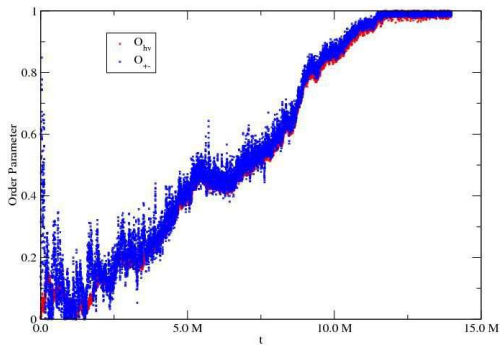} \\
\caption{\label{fig:order400} Order parameters as a function of time in MCS for a quench from the saturated magnetic state, $M(0)=1 $, to $T=4.00$.  }
\end{center}
\end{figure}

\begin{figure}
\begin{center}
\includegraphics[height=45mm]{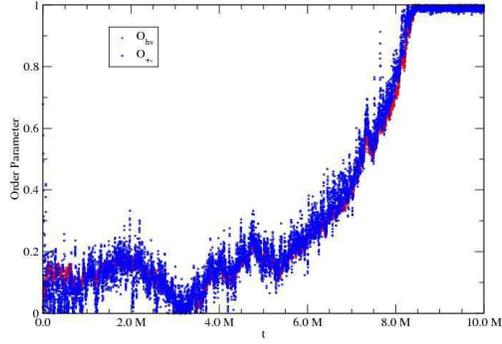} \\
\caption{\label{fig:order425} Order parameters as a function of time in MCS for a quench
 from the saturated magnetic state, $M(0)=1$, to $T=4.25$.  }
\end{center}
\end{figure}

\begin{figure}
\begin{center}
\includegraphics[height=45mm]{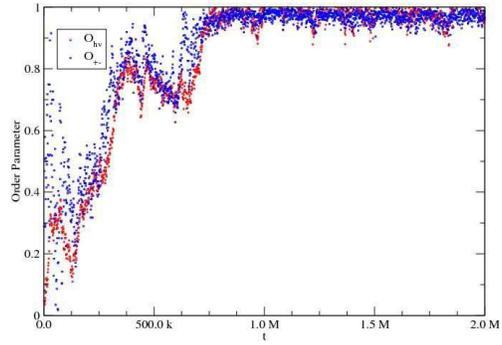} \\
\caption{\label{fig:order490} Order parameters as a function of time in MCS for a quench
 from the saturated magnetic state, $M(0)=1$, to $T=4.90$.  }
\end{center}
\end{figure}

\begin{figure}
\begin{center}
\includegraphics[height=45mm]{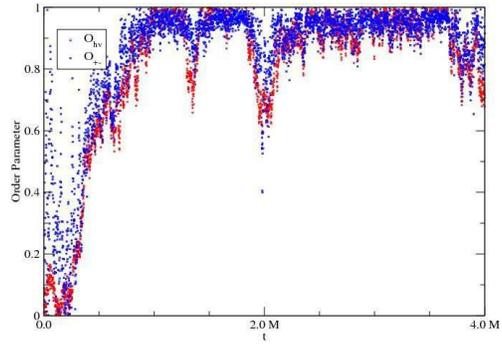} \\
\caption{\label{fig:order500} Order parameters as a function of time in MCS for a quench
 from the saturated magnetic state, $M(0)=1$, to $T=5.00$.  }
\end{center}
\end{figure}

In Figures \ref{fig:order400},  \ref{fig:order425}, \ref{fig:order490},
 and \ref{fig:order500}  we show 
both $\left|O_{hv}\right| $ and $\left|O_{\pm}\right|$ at $T= 4.00, 4.25, 4.90 $ 
\, and \, $5.00$ respectively from a
single quench at each temperature. It is important to note that these
are from single simulations with no averaging and therefore we are
generally looking at qualitative features.  Figures
\ref{fig:order400} to \ref{fig:order500} are for temperatures below or at 
$T_{c}$ and show very similar features, although they have different
times scales.  


\begin{figure}
\begin{center}
\includegraphics[height=45mm]{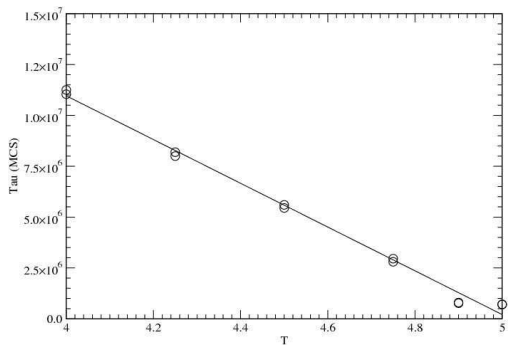} \\
\caption{\label{fig:tauvsT} Estimate of the time $(\tau) $ in MCS for the order
parameter, $|O_{hv}|$, to reach 99\% of its equilibrium value as a function of the
temperature of the simulation. The line is a least squares fit to the points shown and serves only as a guide to the eye.}
\end{center}
\end{figure}

In each simulation at each temperature there are three
stages or regimes.   The order parameter starts at
zero, as it should be in the initial, ferromagnetic state, and
remains near zero at early times.   For $T=4.00$ and $4.25$ this stage
can be up to 2 million to 4 million MCS, while at $T=4.90$ and
$T=5.00$ it is considerably shorter; on the order of 100,000 MCS.
After this stage the order parameter begins to grow 
over a time period which is again temperature dependent, until it saturates 
at its equilibrium value.  At this point the system has reached equilibrium and fluctuates about its
equilibrium value.  We can estimate when the system reaches equilibrium by
measuring the time it takes the order parameter, $O_{hv}$,  to first reach 99\% of its equilibrium value. For the twelve runs, two at each of six temperatures below $T_{c}$,  we
can plot this value versus temperature as shown in Figure \ref{fig:tauvsT}.
One will note that below $T$ for 
all the temperatures considered the system does not ``freeze'' into a 
metastable tetragonal state as suggested by Mu and Ma, but rather the system 
continues to evolve slowly towards the equilibrium state, with a 
characteristic time that grows as a function of the depth of the quench. Once the system has
reached equilibrium the order parameter fluctuates about its equilibrium mean value. In other words
it does not appear that this characteristic time will diverge at finite temperature, so we find no 
evidence of a glass-like state.  Above $T_{c}$ both order parameters show the 
system does not possess any significant orientational order as one would expect.

\section{ Cluster properties \label{sec:count} }

Now we wish to determine if it is possible to qualitatively explain 
the non-equilibrium relaxation of the magnetization and the order parameter using the
properties of the clusters of stripes identified using the  methods introduced earlier.  
We will show configurations from a simulation at $T=4.00$ for
illustrative purposes and discuss properties at all simulation temperatures.

\begin{figure}[htb]
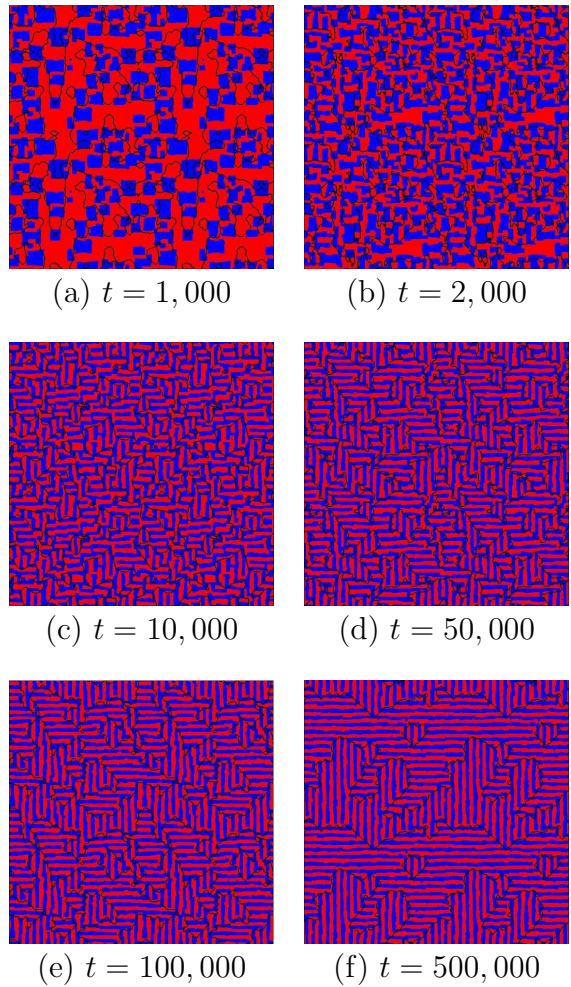

\begin{center}
\begin{tabular}{c c }
\includegraphics[height=35mm]{time=0000001000__4.000-SpinPhase.eps3} &
\includegraphics[height=35mm]{time=0000002000__4.000-SpinPhase.eps3} \\
\large{(a) $t=1,000$} & 
\large{(b) $t=2,000$}\\\\ 
\includegraphics[height=35mm]{time=0000010000__4.000-SpinPhase.eps3} &
\includegraphics[height=35mm]{time=0000050000__4.000-SpinPhase.eps3} \\
\large{(c) $t=10,000$} & 
\large{(d) $t=50,000$}\\\\ 
\includegraphics[height=35mm]{time=0000100000__4.000-SpinPhase.eps3} &
\includegraphics[height=35mm]{time=0000500000__4.000-SpinPhase.eps3} \\
\large{(e) $t=100,000$} & 
\large{(f) $t=500,000$}
\end{tabular}
\caption{ \label{fig:early400} Configurations for a quench from a saturated magnetic state to $T/g=4.00$ at various times from one simulation. }
\end{center}
\end{figure}

Figure \ref{fig:early400} shows the spin configuration from one
simulated quench at $T=4.00$ at early times, t:(a) $ t=1000$,(b)
$t=2000$,(c) $t=10,000$ and (d) $t=100,000$. 
One can clearly see early dynamics similar to those described by
Bromley et al.,  Cannas et al. and  Mu and Ma.
Initially there is nucleation of small spin clusters, which
grow and transform into the labyrinth stripe patterns which are
clearly evident at $T=4.00$ by $t=10,000$ in  Figure \ref{fig:early400}c. 
This is consistent both with the magnetization we have measured and 
the results of the earlier studies. It is also consistent with the order parameter at early 
times, having a value very close to 
zero, as the stripes are just beginning to manifest themselves and hence there is no real orientational
order.  By $t=10,000 MCS$ our cluster identification scheme is able to partition
the system in to domains based on the stripe orientation and these clusters of stripes grow and 
coalesce with time until approximately  $t=100,000$ MCS at $T=4.00$.  At that point the size of the 
clusters of stripes are comparable to our system size.  This is typically the regime that has been considered previously.  

\begin{figure}
\begin{center}
\includegraphics[height=45mm]{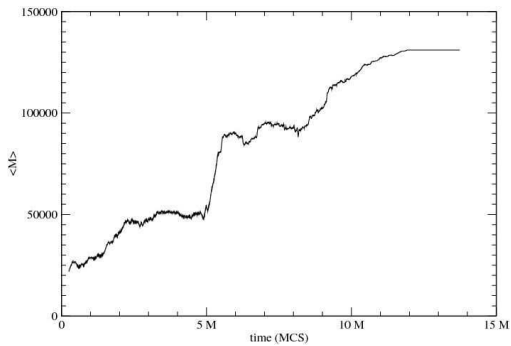} \\
\caption{The average size (mass) of a cluster of stripes as a function of time in MCS at $T=4.00$.
 \label{fig:averagemass} }
\end{center}
\end{figure}

\begin{figure}[htb]
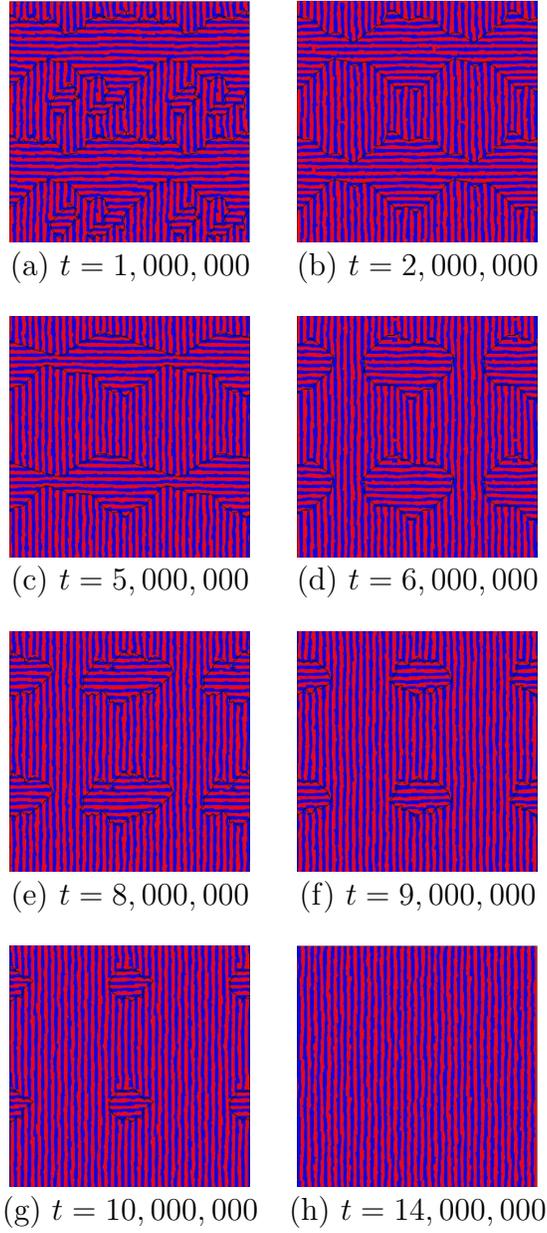

\begin{center}
\begin{tabular}{c c }
\includegraphics[height=32mm]{time=0001000000__4.000-SpinPhase.eps3} &
\includegraphics[height=32mm]{time=0003000000__4.000-SpinPhase.eps3} \\
\large{(a) $t=1,000,000$} & 
\large{(b) $t=2,000,000$}\\\\ 
\includegraphics[height=32mm]{time=0005000000__4.000-SpinPhase.eps3} &
\includegraphics[height=32mm]{time=0006000000__4.000-SpinPhase.eps3} \\
\large{(c) $t=5,000,000$} & 
\large{(d) $t=6,000,000$}\\\\ 
\includegraphics[height=32mm]{time=0008000000__4.000-SpinPhase.eps3} &
\includegraphics[height=32mm]{time=0009000000__4.000-SpinPhase.eps3} \\
\large{(e) $t=8,000,000$} & 
\large{(f) $t=9,000,000$}\\\\ 
\includegraphics[height=32mm]{time=0010000000__4.000-SpinPhase.eps3} &
\includegraphics[height=32mm]{time=0014000000__4.000-SpinPhase.eps3} \\
\large{(g) $t=10,000,000$} & 
\large{(h) $t=14,000,000$}\\\\ 
\end{tabular}
\caption{Configurations for a quench from a saturated magnetic state to
$T/g=4.00$.\label{fig:longtime} }
\end{center}
\end{figure}

Figure~\ref{fig:averagemass} shows the average size (mass) of the clusters of stripes in the simulation
as a function of time in MCS from a single simulation at $T=4.00$.  
This average includes both horizontal and vertical stripe clusters.  The different dynamical regimes are clearly seen in this Figure. The initial regime has ended after at most 
$200,000$ MCS.  The intermediate regime, from $t=200,000$ to $t=11,000,000$ occurs when the system has phase separated into a few very large stripe domains.  Spin configurations from this region are shown in Figure \ref{fig:longtime}. 

\begin{figure}
\begin{center}
\includegraphics[height=45mm]{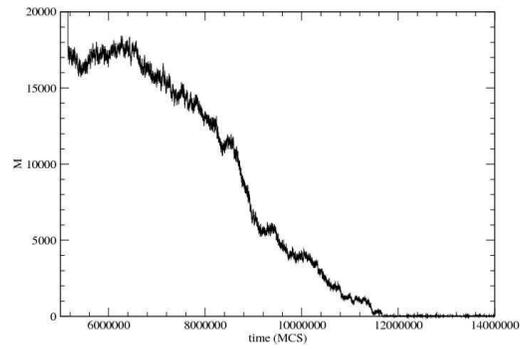} \\
\caption{The size (mass) of the last cluster of the minority stripe phase as a function of time in MCS at $T=4.00$.  In the time frame shown there is only one cluster of the minority stripe phase. \label{fig:bigcluster} }
\end{center}
\end{figure}

The feature in Figure \ref{fig:averagemass} at approximately $t=5 \mbox{ million} $ MCS is  an artifact of the last cluster of horizontal stripes no longer spanning the system, such that instead of counting as 2 clusters in our tiled system it counts as 4 half size clusters.  At that time the system consists of a majority phase with a single stripe domain of the minority stripe phase embedded within it.  The dynamics of the system is now 
defined by the slow evaporation of this domain, which occurs as defects in the stripes are removed from the system.  These defects occur along the boundary of the last cluster of stripes of the minority phase. To show this we measure the mass of the largest cluster of the minority stripe phase as a function of time.  In Figure \ref{fig:bigcluster} we show the mass in the time frame from $t= 5\mbox{ million}$   to $t=14 \mbox{ million}$. Again as this is data from a single run rather than an ensemble average from many runs, one is looking at the qualitative behavior of the system rather than trying to extract the functional form of the decay.

\section{ Summary \label{sec:concl} }
In summary we have shown how to simply and efficiently define clusters of stripes within the dipolar Ising model based on the local stripe orientation and have tracked the growth and decay of such clusters during a quench from a saturated magnetic state to a number of temperatures below the critical temperature.   We have considered the non-equilibrium relaxation by observing the time dependence of both the magnetization and order parameter and by observing the time dependence of these stripe domains.  We have confirmed earlier results concerning the early and intermediate stages of this relaxation and have provided the longest simulations to date to allow the study of the late stage relaxation.  Our results indicate that the late stage relaxation is dominated by the very slow evaporation of clusters of the minority stripe phase.  In all cases we have found that the system relaxes into the smectic, striped  ground state with no indications of freezing into a spin-glass-like state.

This work was supported by the Natural Sciences and Engineering
Council of Canada (NSERC).  The authors would like to acknowledge SHARCNET, 
for the provision of the computing resources used in this work.

\bibliography{paper}
\end{document}